\documentclass[reprint,amsmath,amssymb,aps,prl]{revtex4-2}
\usepackage{graphicx}
\usepackage{dcolumn}% Align table columns on decimal point
\usepackage{bm}% bold math
\usepackage{blindtext}
\usepackage{xcolor}
\UseRawInputEncoding

\newif\ifarxiv
\arxivtrue % Uncomment to not roll in SM.

\begin{document}

\preprint{APS/123-QED}

\title{Edge-sharing quasi-one-dimensional cuprate fragments in optimally substituted Cu/Pb apatite}

\author{Katherine Inzani$^{1}$}
\email{katherine.inzani1@nottingham.ac.uk}
\affiliation{$^{1}$School of Chemistry, University of Nottingham, University Park, Nottingham NG7 2RD, United Kingdom}

\author{John Vinson$^{2}$}
\email{john.vinson@nist.gov}
\affiliation{$^{2}$Material Measurement Laboratory, National Institute of Standards and Technology, Gaithersburg, Maryland, 20899, USA}

\author{Sin\'{e}ad M. Griffin$^{3,4}$}
\email{sgriffin@lbl.gov}
\affiliation{$^{3}$Materials Sciences Division, Lawrence Berkeley National Laboratory, Berkeley, California, 94720, USA}
\affiliation{$^{4}$Molecular Foundry Division, Lawrence Berkeley National Laboratory, Berkeley, California, 94720, USA}

\date{\today}

\begin{abstract}
The flurry of theoretical and experimental studies following the report of room-temperature superconductivity at ambient pressure in Cu-substituted lead apatite Cu$_x$Pb$_{10-x}$(PO$_4$)$_6$O (`LK99') have explored whether and how this system might host strongly correlated physics including superconductivity. While first-principles calculations at low doping ($x\approx1$) have indicated a Cu-$d^{9}$ configuration coordinated with oxygen giving rise to isolated, correlated bands, its other structural, electronic, and magnetic properties diverge significantly from those of other known cuprate systems. Here we find that higher densities of ordered Cu substitutions can result in the formation of contiguous edge-sharing Cu-O chains, akin to those found in some members of the cuprate superconductor family. Interestingly, while such quasi-one-dimensional edge-sharing chains are typically ferromagnetically coupled along the chain, we find an antiferromagnetic ground-state magnetic order for our cuprate fragments which is in proximity to a ferromagnetic quantum critical point.  This is the consequence of the elongated Cu-Cu distance in Cu-substituted apatite that leads to larger Cu-O-Cu angles supporting antiferromagnetism, which we demonstrate to be controllable by strain. Finally, our electronic structure calculations confirm the low-dimensional nature of the system and show that the bandwidth is driven by the Cu-O plaquette connectivity, resulting in an intermediate correlated regime. 

\end{abstract}

\maketitle

%\section{Introduction}
\paragraph*{Introduction ---} 
Cuprate systems exhibit a rich playground of strongly-correlated physics encompassing phenomena from high-temperature superconductivity to frustrated, low dimensional quantum spin systems~\cite{Bednorz_et_al:1986, Keimer_et_al:2015}. Despite this diversity of emergent phenomena, several key features are shared across these cuprate systems. The stoichiometric compounds are characterized as Mott or charge-transfer insulators with $S=\frac{1}{2}$ moments on Cu ions stemming from a Cu-$d^{9}$ configuration. These Cu ions are coordinated with oxygen to form CuO$_4$ square planes that are typically antiferromagnetically coupled. Electron- or hole-doping these antiferromagnetic insulators can suppress the magnetic order and result in superconductivity. It is these superexchange-driven correlations that are suggested to give rise to potential $d$-wave superconductivity~\cite{Scalapino_et_al:1986, Anderson:1987}. 

The fundamental building blocks of the cuprates are square planar coordinated CuO$_4$ that form either corner- or edge-sharing networks. Corner-sharing chains, found in cuprate superconductors and spin ladders, exhibit strong nearest-neighbor exchange interactions (100~meV to 160~meV) facilitated through an $\approx180^\circ$ Cu-O-Cu pathway. Conversely, edge-sharing chains have weaker nearest-neighbor exchange interactions due to their $\approx 90^\circ$ Cu-O-Cu pathway, resulting in a range of reported ferromagnetic and antiferromagnetic behaviors dependent on the specific Cu-O-Cu bond angle. 

Reducing the dimensionality of these $S=\frac{1}{2}$ systems, from the well-known 2D antiferromagnetic Heisenberg model to 1D, results in a significant enhancement of quantum fluctuations. In fact, Bethe predicted such strong quantum fluctuations would prevent long-range magnetic order from forming in  the 1D $S=\frac{1}{2}$ antiferromagnetic Heisenberg chain \cite{bethe_zur_1931}. The diversity in structural configurations of cuprates, influenced by factors such as connectivity, orbital overlap, and dimensionality, highlights the complexity of the physical landscape inherent in cuprates, leading to a wide spectrum of observed properties in these compounds. Recently, superconductivity was reported in the infinite-chain cuprate Sr$_x$Ca$_{1-x}$CuO$_2$ with a T$_C$ of 90~K~\cite{Rajak_et_al:2023}, a material in which the 1D chains have features of both edge- and corner-sharing Cu-O-Cu pathways.  However, the limited number of 1D cuprate systems, and the challenge of doping them, have hindered progress in exploring the phase diagram of 1D cuprate chains and possible routes to superconductivity in these systems~\cite{Chen_et_al:2021}.

\begin{figure*}
\includegraphics[width=0.85\textwidth]{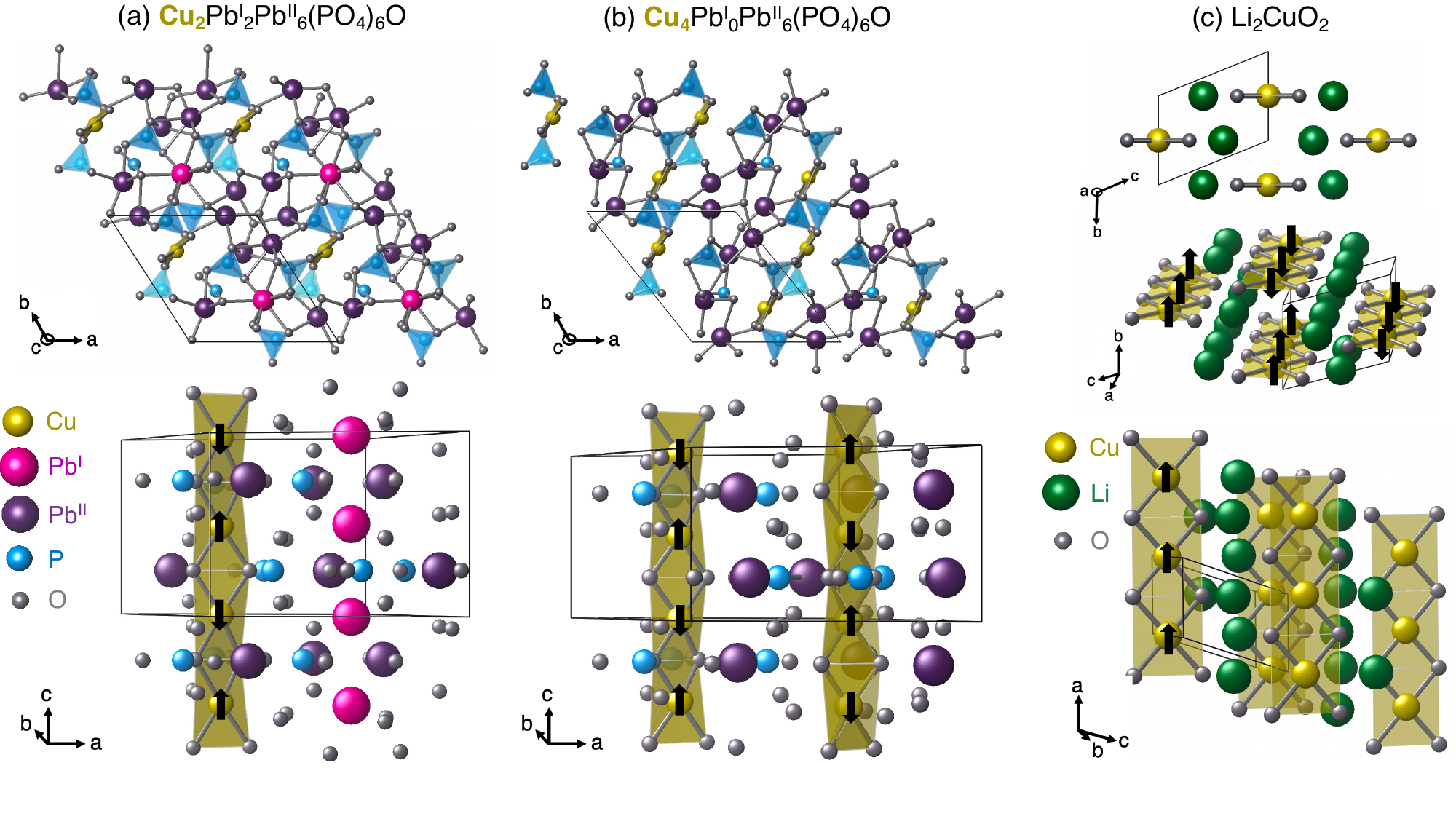}
\caption{Crystal structures with the primitive unit cells of the non-magnetic cell shown by the box. The arrows depict the calculated ground-state magnetic order.  (a) Cu$_2$Pb$^\mathrm{I}_{2}$Pb$^\mathrm{II}_6$(PO$_{4}$)$_6$O with (half) two Pb$^\mathrm{I}$ sites substituted with Cu. (b) Cu$_4$Pb$^\mathrm{I}_{0}$Pb$^\mathrm{II}_6$(PO$_{4}$)$_6$O with all four Pb$^\mathrm{I}$ sites substituted with Cu. (c) Orthorhombic Li$_2$CuO$_2$.} 
\label{structures}
\end{figure*}

The recent report of possible room temperature superconductivity at ambient pressures in Cu-substituted Pb apatite (also known as `LK99')~\cite{RoomT,RoomT2} has focused attention on their reported  CuPb$_9$(PO$_4$)$_6$O structure. Initial calculations using Density Functional Theory (DFT) showed that Cu substitution on the trigonal-prismatically coordinated Pb site results in a $d^9$ electronic configuration with a narrow, isolated, half-filled Cu-$d$ band in its paramagnetic phase~\cite{Griffin:2023}, whose main features can be described by a minimal two-band model~\cite{Tavakol/Scaffidi:2023}. The initial DFT work, and subsequent theoretical studies predicted it to have a frustrated antiferromagnetic ground state at the DFTlevel~\cite{Dessau:2023}, however with very weak coupling. Appropriately incorporating strong correlations through Dynamical Mean-Field Theory, showed it to lie in an ultra-correlated regime ($U/W \approx 30-50$) which discounted known mechanisms for electronically-driven superconductivity~\cite{lk99-Werner:2023}. This motivates us to consider how the bandwidth can be increased to a more moderately correlated regime via chemical and physical modifications to the Cu-substituted apatite framework. In this paper, we perform DFT calculations of Cu-substituted lead apatite in higher Cu concentrations, namely Cu$_x$Pb$_{10-x}$(PO$_4$)$_6$O (for $x=2, 4$), identifying edge-shared square planar arrays of CuO$_4$ plaquettes. We examine the magnetic and electronic phase diagram of this system, comparing its key features with the prototypical examples of cuprates with  (i) edge-sharing motifs, Li$_2$CuO$_2$ and (ii) corner-sharing motifs, La$_2$CuO$_4$.

%\section{Methods}
\paragraph*{Methods ---} 
All of our DFT calculations used the Vienna Ab initio Simulation Package (VASP) \cite{Kresse1993,Kresse1994,Kresse1996,Kresse1996a}, with further details given in the SI. To account for Cu-$d$ localization, we applied a Hubbard-U correction of 8~eV which was benchmarked in recent work with meta-GGA functionals and calculation of self-consistent U~\cite{site-selectivity-paper}.  Electronic structure plots were generated using the sumo software package~\cite{sumo}, and Fermi surfaces were generated using iFermi~\cite{iFermi}. We also benchmarked our calculations by repeating the electronic structure calculation for a representative system with the meta-GGA r2SCAN functional \cite{r2scan}, finding qualitatively similar results (see SI).

%\section{Results}
%\subsection{Structural Properties}
\paragraph*{Structural Properties ---} 
We consider the family of apatites given by Cu$_x$Pb$^\mathrm{I}_{4-x}$Pb$^\mathrm{II}_6$(PO$_{4-y}$)$_6$\textit{X}$_2$, where \textit{X} = O$^{2-}$, S$^{2-}$, F$^{-}$, Cl$^{-}$, etc. Here, we limit the Cu substitution to the Pb$^\mathrm{I}$ site and \textit{X} = O$^{2-}$, and  proper charge is maintained by $y=1/6$, reflecting evidence that two O$^{2-}$ ions in the channel are balanced with O vacancies elsewhere~\cite{Keimer:2023}. 
%Some work has indicated important symmetry stuff, but we simply to consider only something
Our choice of only substituting Cu on site I is motivated by the markedly different behavior of the Cu-$d$ states on site I compared to site II. At $x=1$ substitution and ambient pressure synthesis, neither site I and II occupancies are strongly favored \cite{Keimer:2023}. However, Ca-doped Pb apatite shows a strong affinity for ordered substitution on site I when Ca:Pb ratio is 2:3 due to the mismatch in cation size, known as the hedyphane sub-group of apatites, facilitating a smaller-than-trend unit cell compared to the otherwise solid-solution behavior \cite{Rouse:1984,Bigi:1989,Bigi:1991}. 
Theoretical investigations using DFT support this approach, indicating that site I is preferred for $x=2$ substitution (Cu$_2$Pb$_8$(PO$_4$)$_6$) \cite{Wolverton:2023}, and, for $x=1$ substitution, that site I is preferred at high pressures \cite{site-selectivity-paper,Ogawa:2023}. It is possible that site selectivity can also be achieved through changes to the channel ion $X$ \cite{wang2024possible,site-selectivity-paper}. We therefore calculate Cu substitution on site I for both $x=2$ `\textit{Cu-2}' and $x=4$ `\textit{Cu-4}'.
%, as depicted in Fig.~\ref{structures}, noting that the original papers considered $x=1$.

The optimized structures for \textit{Cu-2} and \textit{Cu-4} are shown in Fig.~\ref{structures} where both include an oxygen vacancy (see SI for structure files). The inclusion of oxygen vacancies seems crucial for the formation of Cu-O chains -- without oxygen vacancies (and with only a single channel oxygen) our calculations indicate that contiguous plaquettes do not form (see SI). This suggests that the formation of continguous plaquettes is sensitive to Cu site-selectivity, Cu clustering, and oxygen vacancies. We find a significant lattice collapse for both cases as a result of the substitution of the smaller Cu$^{2+}$ ion (0.73~\AA) for Pb$^{2+}$ site (1.19~\AA) -- the out-of-plane lattice parameters reduce to $c=6.95$~\AA\ and $c=5.99$~\AA\ for \textit{Cu-2} and \textit{Cu-4}, respectively. Cu substitution on the Pb$^\mathrm{I}$ site distorts the original trigonal antiprism oxygen coordination to now being square planar, resulting in a new monoclinic space group of \textit{Pm} (\#6). 
Looking closely at the new Cu coordination, we find that the $c$-lattice reduction facilitates the formation of contiguous edge-sharing CuO$_4$ plaquettes. Such edge-sharing cuprate fragments are less commonly found than their corner-sharing counterparts, and are typically seen in high-pressure phases such as SrCuO$_2$ and in orthorhombic Li$_2$CuO$_2$ (Fig.~\ref{structures}(c)).

\begin{figure}
\includegraphics[width=0.48\textwidth]{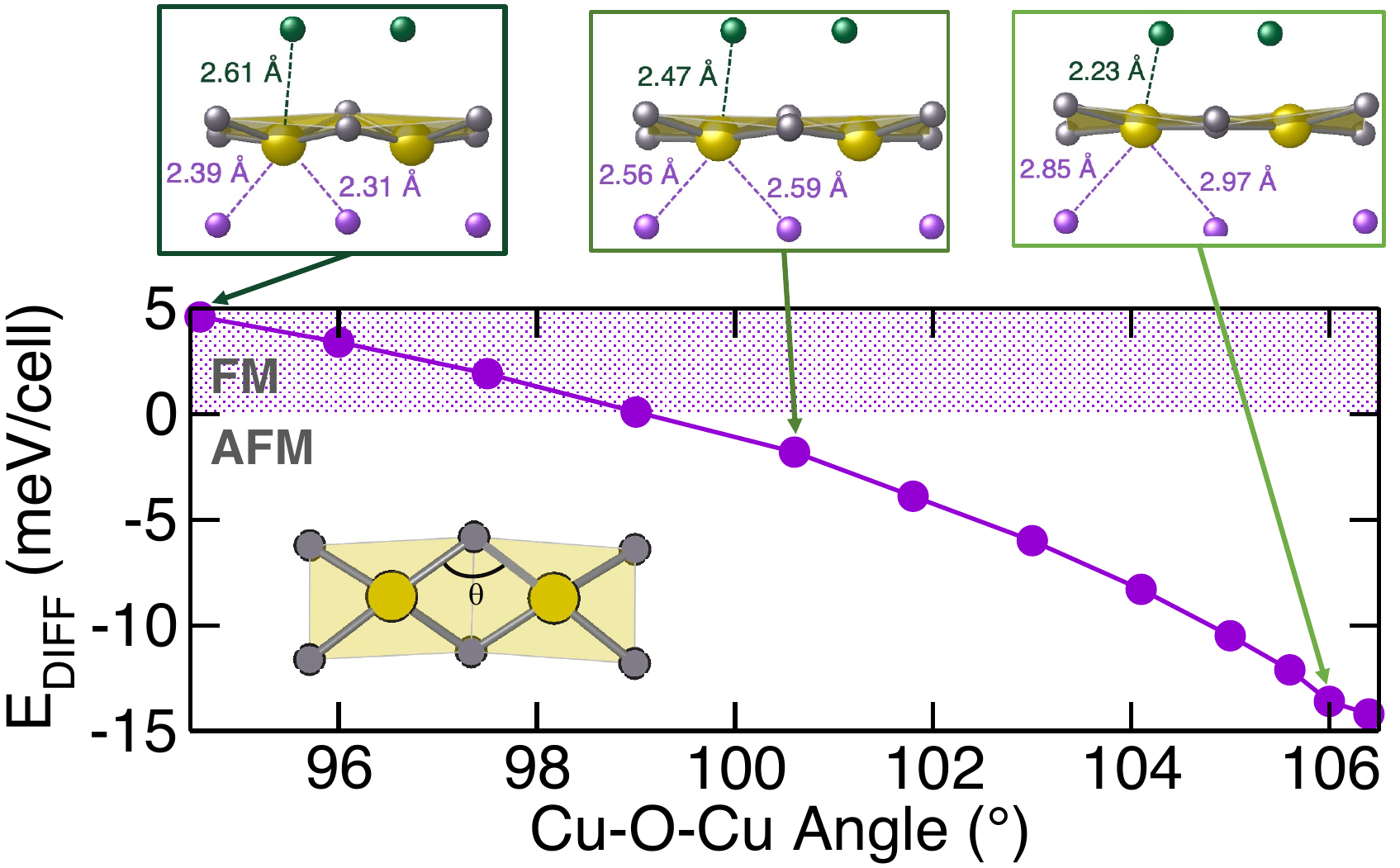}
\caption{Calculated energy difference between ferromagnetic and antiferromagnetic coupling along the Cu-O chains for Cu$_2$Pb$^\mathrm{I}_{2}$Pb$^\mathrm{II}_6$(PO$_{4}$)$_6$O as a function of average Cu-O-Cu bond angle when uniaxial out-of-plane strain is applied. Negative values correspond to an antiferromagnetic ground state. The top panel shows the evolution of apical oxygen bond lengths for a variety of configurations corresponding to 94.6$^\circ$, 101.8$^\circ$ and 106.4$^\circ$.} 
\label{angle-magnetism}
\end{figure}

The reduction in $c$ lattice length is likely necessary to support the formation of edge-sharing CuO$_4$ plaquettes. Known edge-sharing cuprates have a Cu-Cu separation that ranges from 2.84~\AA{} in Li$_2$CuO$_2$~\cite{Li2CuO2-crystal} to 2.94~\AA{} in CuGeO$_3$~\cite{CuGeO3-crystal}, as compared to our results of 3.46/3.48~\AA{} in \textit{Cu-2} and 2.98/3.01~\AA{} in \textit{Cu-4}.
We find that the Cu-O bond distances range from 2.07~\AA{} to 2.21~\AA{} in \textit{Cu-2} to 1.95~\AA{} to 2.01~\AA{} in \textit{Cu-4}, compared to 1.95~\AA{} for both Li$_2$CuO$_2$~\cite{Li2CuO2-crystal} and CuGeO$_3$~\cite{CuGeO3-crystal}. 
This variation in Cu-O bond lengths of the apatites is due in part to the fact that the CuO$_4$ plaquettes in \textit{Cu-2} and \textit{Cu-4} are puckered out-of-phase where the coordinating oxygen are staggered in the plane perpendicular to the chain (see Fig.~\ref{angle-magnetism}). This puckering is also observed in some high-pressure phases of cuprates such as incommensurate Sr$_{0.73}$CuO$_2$~\cite{Schwer_et_al:1997}.

\begin{figure}
\includegraphics[width=0.48\textwidth]{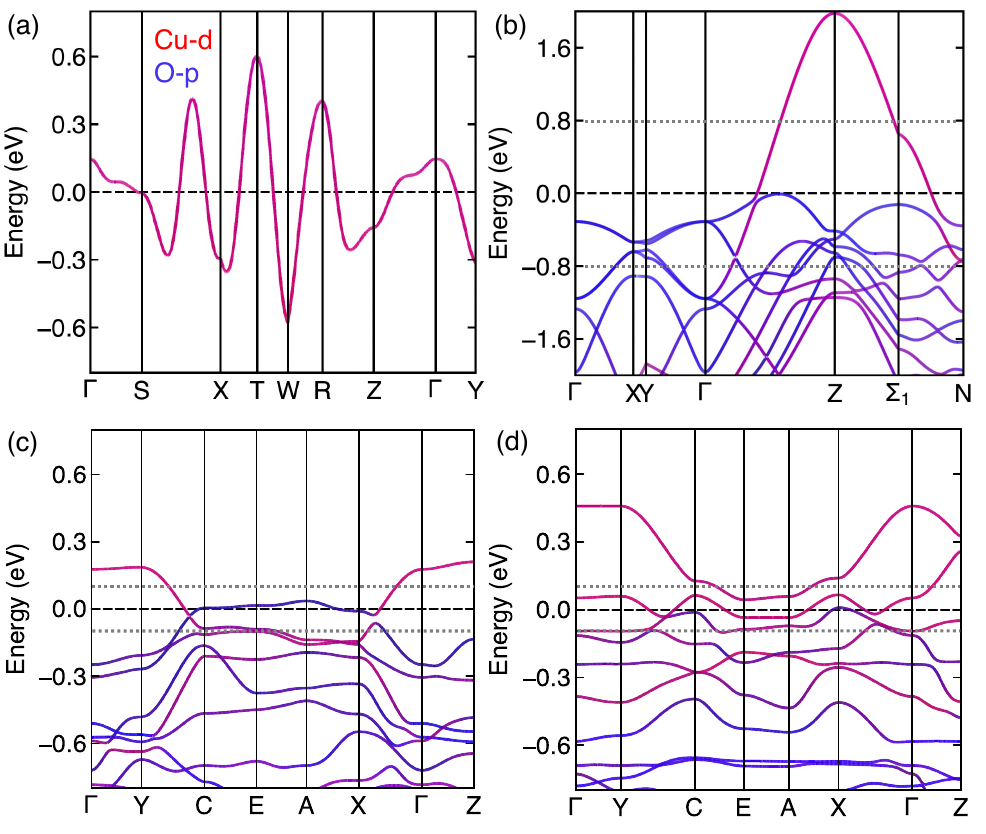}
\caption{Calculated non-spin-polarized electronic band structures for (a) Li$_2$CuO$_2$, and (b) La$_2$CuO$_4$, (c) \textit{Cu-2} Cu$_2$Pb$_{8}$(PO$_{4}$)$_{6}$O, and (d)  \textit{Cu-4} Cu$_4$Pb$_{6}$(PO$_{4}$)$_{6}$O. The red and blue show the orbital projections of the Cu-$d$ and O-$p$ states respectively. The Fermi level is set to 0 eV in all plots and is marked by the dashed line. Note that the energy range for (b) is greater than the other three -- for comparison the energy range of (a), (c) and (d) is marked by the dotted lines. $E_{\textrm{F}}\pm0.1$~eV is also marked by dotted lines on (c) and (d).} 
\label{bands-pm}
\end{figure}

%\subsection{Electronic Structure}
\paragraph*{Electronic Structure ---}

We first present the calculated electronic structures without including spin-polarization which is more relevant for temperatures above the magnetic Curie/N\'{e}el temperature (spin-polarized calculations are given in the SI). These are plotted in Fig.~\ref{bands-pm} for the two Cu-ordered cases \textit{Cu-2} and \textit{Cu-4}, and for the known cuprates Li$_2$CuO$_2$ and La$_2$CuO$_4$ for a region around the Fermi level. As expected from Cu-$d^9$ in a square-planar crystal field environment in all structures, we find half-filled bands at the Fermi level dominated by the highest-energy Cu-$d$ orbital. We note that the nonmagnetic primitive unit cell of Li$_2$CuO$_2$ contains only one Cu ion whereas that of \textit{Cu-2} and \textit{Cu-4} contain two and four respectively, resulting in the half-filling of two and four bands at the Fermi level.

However, beyond this, the electronic structures of these cuprate compounds diverge. For Li$_2$CuO$_2$, the splitting of the hole orbital is much greater than that of La$_2$CuO$_4$, \textit{Cu-2}, or \textit{Cu-4} resulting in an isolated band separated by over 1 eV from the rest of the valence manifold. This is in contrast to the other three compounds whose hole band hybridizes with the top of the valence band. The bandwidth of Li$_2$CuO$_2$ is $\approx$1.1~eV, as a result of  hopping along edge-sharing Cu-O chains with a Cu-Cu separation of 2.83~\AA. In contrast, the Cu-Cu separation for \textit{Cu-2} and \textit{Cu-4} are 3.46/3.48~\AA\ and 2.98/3.01~\AA\, resulting in reduced bandwidths of approximately 0.5~eV and 1~eV, respectively. These bandwidths are greatly increased from the $x=1$ case reported previously of around 0.1~eV~\cite{Griffin:2023}, pushing these heavily substituted \textit{Cu-2} and \textit{Cu-4} apatites into a more intermediately correlated regime where $U/W \approx 1-10$ as desired. The corner-shared La$_2$CuO$_4$ has a bandwidth of over 3~eV, which at first glance might be surprising given its Cu-Cu distance of 3.95~\AA, the largest of all of the cases considered here. The difference is in the Cu-O-Cu bond angle. For the case of the previously discussed edge-shared Cu-O plaquettes, the Cu-$d$ and O-$p$ are  orthogonal with a Cu-O-Cu bond angle of 90$^{\circ}$, resulting in a strongly suppressed hopping. However, the corner-shared systems like La$_2$CuO$_4$ have significant Cu-$d_{x^2-y^2}$ hybridization with O-$p$ resulting in increased bandwidth. 

For corner-sharing square plaquettes, like those in La$_2$CuO$_4$, the orbital order is dominated by crystal-field splitting of the square planar array which results in the hole orbital being primarily Cu-$d_{x^2-y^2}$ with some O-$p$ hybridization (see SI. Fig.~\ref{pdos-comb}(b)) as reported previously~\cite{Schluter_et_al:1988}. However, for the edge-sharing CuO$_2$ plaquettes, metal-metal bonding results in a redistribution of the orbitals such that the in-plane $\sigma$-type Cu-$d$ orbital is half-filled. This corresponds to the orbital that is parallel to the Cu-O-Cu chain, and perpendicular to the interchain direction~\cite{Huang_et_al:2011}. For the case of Li$_2$CuO$_2$, the hole band therefore has Cu-$d_{xz}$ character with significant hybridization found with O-p$_z$, as previously reported~\cite{Weht/Pickett:1998}. Similarly for both \textit{Cu-2} and \textit{Cu-4}, the chain runs in the out-of-plane direction, resulting in a hole band of Cu-$d_{xz}$ with some O-$p_{z}$ (see SI. Fig.~\ref{pdos-comb}(c,d)).

\begin{figure}
\includegraphics[width=0.48\textwidth]{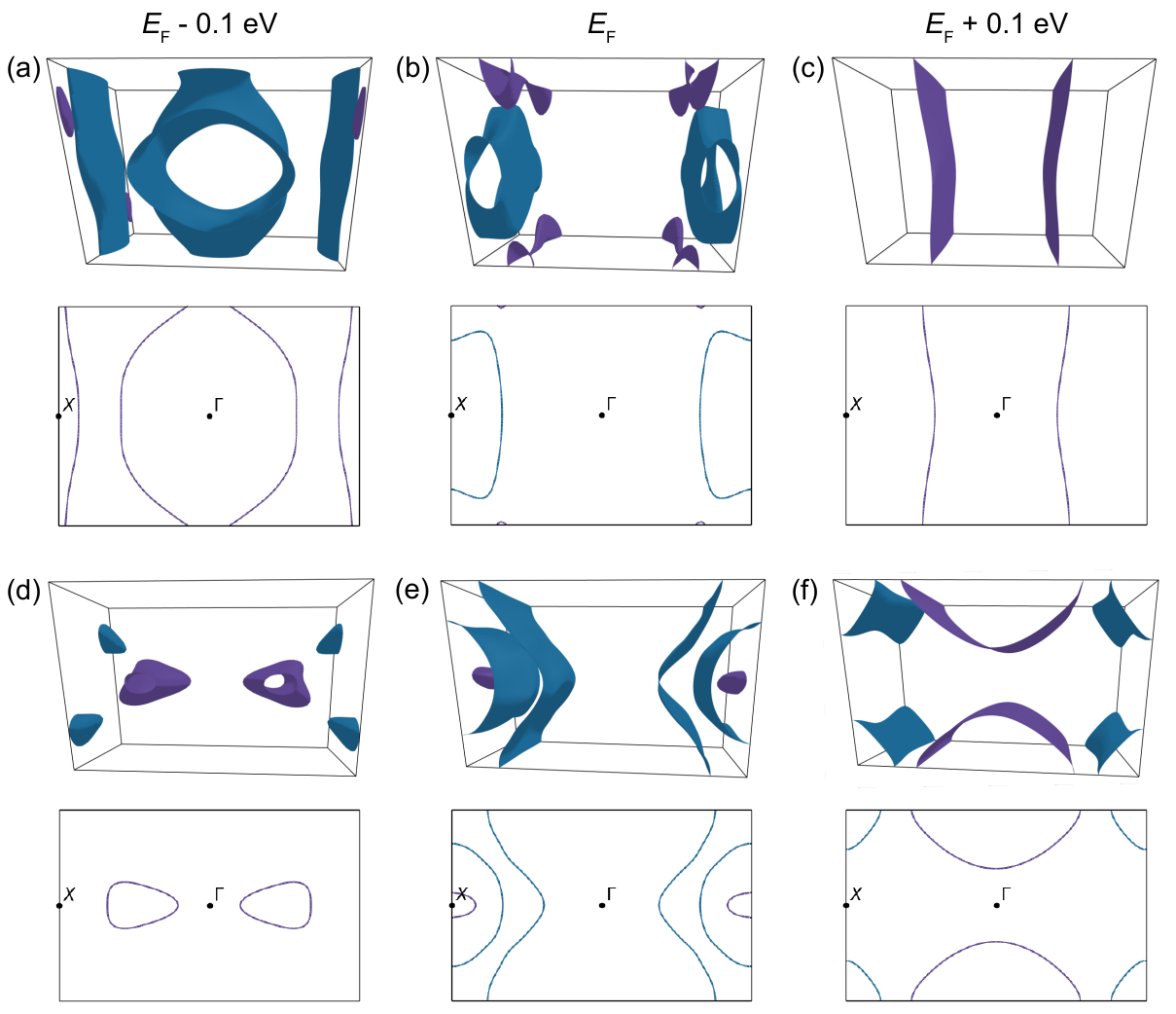}
\caption{Calculated non-spin-polarized Fermi surfaces for \textit{Cu-2} at (a) $E_{\textrm{F}}-0.1$~eV, (b) $E_{\textrm{F}}$, and (c) $E_{\textrm{F}}+0.1$~eV, and for \textit{Cu-4} in (d), (e) and (f) respectively.} 
\label{fermi}
\end{figure}

In Fig.~\ref{fermi}, we plot the non-spin-polarized Fermi surfaces for \textit{Cu-2} and \textit{Cu-4} for a range of values of the Fermi level. Interestingly, for both cases we find nested Fermi surfaces and a strong dependence of the dimension and nesting on the position of the Fermi level. For instance,  \textit{Cu-4} transitions from being 3D to more 2D in going from 0.1 eV below the Fermi level to the Fermi level. The Fermi surface character also changes from comprised of concentric cylinders surrounding the center of the BZ edge ($X$ point) at the Fermi level to two separate cylinders centered on the BZ edge and corner at 0.1 eV above the Fermi level. This sensitivity of the details of the Fermi surface with doping also suggests strong potential for spin frustration and fluctuations in these systems.

%\subsection{Magnetic Properties}
\paragraph*{Magnetic Properties ---} 

In edge-sharing cuprates, the nature of frustrated interactions is deeply intertwined with the linkage type of CuO$_4$ plaquettes. With a Cu-O-Cu angle adjusting from 180$^\circ$ in corner sharing to 90$^\circ$  in edge sharing, the nature of the exchange interaction dramatically changes. According to the Goodenough-Kanamori-Anderson rule, the Cu-Cu nearest-neighbor exchange interaction is strong and antiferromagnetic across a 180$^\circ$ Cu-O-Cu bond, while a 90$^\circ$ bond results in a weaker and ferromagnetic exchange. This implies that in edge-sharing complexes, the antiferromagnetic superexchange interaction is diminished due to the orthogonality of Cu-$d$ and O-$p$ orbitals. 
%However, an effective exchange interaction is still plausible through the Cu-O-O-Cu path between the next-nearest neighbor (NNN) Cu ions, with the contribution of the NNN exchange interaction being antiferromagnetic. 
However, an effective exchange interaction is still plausible through the Cu-O-O-Cu path between the next-nearest neighbor Cu ions, with this contribution being antiferromagnetic. 
This significant influence of the next-nearest neighbors leads to magnetic frustration in edge-sharing systems. This then results in diverse ground states dependent on the ratios of these competing exchange interactions and the corresponding dependence on the CuO$_4$ connectivity.

We calculated that the ground-state magnetic order of the \textit{Cu-4} structure in the collinear limit is G-type  with antiferromagnetic interactions both along the chains and between the chains. We find this to be 0.9~meV/Cu (5.3~meV/Cu) more favorable than ferromagnetic interactions along the chain (and antiferromagnetic between chains), and 1.1~meV/Cu (5.0~meV/Cu) more favorable than antiferromagnetic along chains that are ferromagnetically coupled, indicating strong frustration. The values in parenthesis are those calculated using the r2SCAN functional, which shows an enhancement of exchange interactions, but maintains the ground-state AFM order. This AFM coupling along the chain can be rationalized by the Goodenough-Kannamori rules that states that in going from a Cu-O-Cu bond angle of 180$^\circ$ to 90$^\circ$, the exchange interaction changes from antiferromagnetic to ferromagnetic. In our case the Cu-O-Cu bond angles vary between 96$^\circ$ and 102$^\circ$, consistent with our DFT-calculated antiferromagnetic spin chain order.

For \textit{Cu-2}, we also find a G-type AFM ground state with AFM coupling both along the chain and between chains. This is only 8~$\mu$eV/Cu (21~$\mu$eV/Cu for r2SCAN) more favorable than AFM coupling along the chain and FM coupling between the chains, as expected from the larger chain separation. For the case of FM coupling along chains that are AFM coupled to each other, we find it to be 7.1~meV/Cu (60~meV/Cu for r2SCAN) higher in energy than both AFM coupled. Like the \textit{Cu-4} case, we find AFM exchange along the chains, independent of functional, however  for \textit{Cu-2}, AFM further stabilized. This is due to the larger $c$ lattice parameter for \textit{Cu-2} which results in much larger Cu-O-Cu bond angles ranging from 106$^\circ$ and 110$^\circ$. Given the strong dependence of exchange interactions on the out-of-plane lattice parameter, $c$, which in turn modifies the Cu-O-Cu bond angle, we next explicitly calculate how uniaxial strain along $c$ can be used to control exchange along the chains. Given the very weak interchain coupling for \textit{Cu-2}, we neglect it for the strain calculations, keeping it fixed as FM. In Fig.~\ref{angle-magnetism} we plot the calculated energy difference between AFM and FM intrachain coupling for \textit{Cu-2} as a function of the smallest Cu-O-Cu bond angle resulting from adjusting the $c$-lattice parameter (parallel to the chain). The fully relaxed \textit{Cu-2} structure corresponds to the right-most datapoint of 106.4$^\circ$. We confirm the strong dependence of exchange interactions on bond angle, uncovering a transition from AFM to FM via a quantum critical point. 

We next compare these results with other reported edge-sharing cuprate systems, in particular for Li$_2$CuO$_2$ where we have performed additional DFT calculations for direct comparison. We confirm that Li$_2$CuO$_2$ has an overall antiferromagnetic ground state, comprising ferromagnetic interactions along the Cu-O chains with the chains then antiferromagnetically coupled to each other. This is consistent with previous DFT calculations and neutron diffraction experiments \cite{Sapina_et_al:1990,Weht/Pickett:1998} where we also find a significant fraction ($\approx$20\%) of the ordered moment to be localized on the O-$p$ states. From our structural optimization we find that the Cu-O-Cu bond angle is 94$^\circ$, putting it closer to the ferromagnetic exchange predictions from the Goodenough-Kannomori rules. Other systems with edge-sharing spin chains also follow this trend where the Cu-O-Cu bond angle dictates the dominant exchange interactions -- this is typified in the example of high-pressure Sr$_{0.73}$CuO$_2$. It adopts an incommensurate structure owing to the mismatch between the Sr ions and Cu-O plaquette periodicities -- this results in a distorted wave of the Cu-O plaquettes with distances ranging from 1.85~\AA\ to 2.05~\AA. Like Li$_2$CuO$_2$, it is also ferromagnetic along the Cu-O chains, which are antiferromagnetically coupled to each other -- the coupling strength along the chain can be modified with the Ca substitution on the Sr site~\cite{Wu_et_al:1999}.  On the other hand, the corner-sharing undoped La$_2$CuO$_4$ adopts a G-type antiferromagnetic order which we calculate to be 68 meV/Cu and 158 meV/Cu more favorable than striped AFM and FM orders respectively, consistent with its 180$^\circ$ Cu-O-Cu bond angle.

%\section{Discussion}
\paragraph*{Discussion ---}

Cu-substituted apatites provide a new system for understanding one-dimensional cuprate chains in a previously inaccessible coupling regime. The large lattice vector (Cu-Cu separation) compared with other edge-sharing cuprates leads to a larger Cu-O-Cu angle. In turn, this enables a stronger effective Cu-Cu nearest-neighbor interaction, stabilizing antiferromagnetic ordering, in contrast to typically ferromagnetic edge-sharing cuprates.  Our calculations show that the {\it Cu-4} structure is very close to a ferromagnetic ground state, indicative of strongly frustrated interactions and proximity to a quantum critical point. As shown in Fig.~\ref{angle-magnetism}, this transition is dependent on the Cu-O-Cu angle, and can be tuned by structural modifications.

The increased band width of the Cu-$d_{xy}$ states at the Fermi level seen in \textit{Cu-2} and \textit{Cu-4} over the previously examined $x=1$ structure is an expected result of halving the Cu-Cu separation along the $c$ direction. While the Cu-Cu separation is large compared to other edge-sharing cuprates, which could be expected to reduce the band width, we find the structures to be in the intermediate coupling regime, favorable for superconductivity. This is in part due to the distortion to larger Cu-O-Cu bond angles which is also seen to stabilize an antiferromagnetic ground state. The oxygen atoms that constitute the plaquettes are also bound to the PO$_4$ units, the size of which, along with the $c$ lattice, constrains the Cu-O bond lengths and the Cu-O-Cu angle. Substitutions for P may allow for a further tuning of the correlation and antiferromagnetic ordering, while substitutions for the channel oxygen with other anions (e.g. S, F, Cl, etc.) could additionally modify the structural properties and site selectivity~\cite{site-selectivity-paper}.
Future work is needed to elucidate the role of phonon coupling and possible symmetry-lowering structural modifications such as the  formation of charge density waves or ferroic symmetry breaking \cite{Griffin:2023,hlinka2023}.

The proposed structure requires substantially higher doping and/or Cu clustering than has been reported experimentally \cite{RoomT,RoomT2,Keimer:2023}, in addition to a high degree of site selectivity to facilitate continguous cuprate chains. This is not insurmountable as both solid-solution and ordered doping of apatite structures are common. Reported synthesis attempts have been carried out under ambient pressure and using oxygen as the channel species \textit{X}. Recent theoretical work has identified that pressure, strain, and channel species call all have a strong effect on the site selectivity of the copper dopants \cite{site-selectivity-paper,Ogawa:2023,Wolverton:2023}. Successfully identifying synthesis routes for stabilizing our edge-sharing cuprate chains through appropriate chemical and physical routes will establish Cu-substituted apatites as an entirely new family for understanding cuprate physics, the 1D Hubbard model, and, potentially, high-T$_C$ superconductivity.

\paragraph*{Acknowledgements ---} 
We are grateful to Alexander Balatsky and Masatoshi Imada for useful feedback. 
%I also thank xyz further comments. 
This work was funded in part by the U.S. Department of Energy, Office of Science, Office of Basic Energy Sciences, Materials Sciences and Engineering Division under Contract No. DE-AC02-05-CH11231 within the Theory of Materials program. Computational resources were provided by the National Energy Research Scientific Computing Center and the Molecular Foundry, DOE Office of Science User Facilities supported by the Office of Science, U.S. Department of Energy under Contract No. DEAC02-05CH11231. The work performed at the Molecular Foundry was supported by the Office of Science, Office of Basic Energy Sciences, of the U.S. Department of Energy under the same contract. K.I. acknowledges support from the EPSRC (EP/W028131/1).

Certain software is identified in in this paper in order to specify the procedure adequately. Such identification is not intended to imply recommendation or endorsement of any product or service by NIST, nor is it intended to imply that it is necessarily the best available for the purpose.

\bibliographystyle{apsrev}
\bibliography{library}

\clearpage

\section{Supplemental Material}
%The following two lines change the figure labels to S1, S2 for the SM
\renewcommand\thefigure{S\arabic{figure}}  
\setcounter{figure}{0} 
%end
\subsection{Calculation Details}

We used the Vienna \textit{Ab initio} Simulation Package (VASP) \cite{Kresse1993,Kresse1994,Kresse1996,Kresse1996a} for all of our Density Functional Theory (DFT) calculations  with projector augmented wave (PAW) pseudopotentials \cite{Blochl1994,KresseA}. We included the Pb 5d$^{10}$6s$^{2}$6p$^{2}$, Cu 3d$^{10}$4s$^{1}$, P 3s$^{2}$3p$^{5}$, O 2s$^{2}$2p$^{4}$, Li 2s$^1$, and La 5s$^{2}$5p$^{6}$5d$^{1}$6s$^{2}$ as valence electrons. For all calculations we used a plane-wave cut-off energy of 700 eV. The Brillouin Zone (BZ) was sampled with a $6\times6\times8$ Gamma-centered k-point grid for structural optimizations and a $10\times10\times12$ Gamma-centered k-point grid for the density of states for the \textit{Cu-2} and  \textit{Cu-4} systems. For Li$_2$CuO$_2$ and La$_2$CuO$_4$, we used a $12\times12\times4$ Monkhorst-Pack k-point grid. We used  the generalized gradient approximation (GGA) based exchange-correlation functional PBEsol \cite{Perdew2008}, and applied a Hubbard-U of 8 eV to the Cu-$d$ states within the Dudarev approach~\cite{Dudarev1998}, unless otherwise stated. Meta-GGA functional calculations used the r2SCAN functional \cite{r2scan}.
The electronic convergence criterion is set to $10^{-6}$ eV and the force convergence criterion is set to 0.01 eV / \AA{}. 

\subsection{Choice of Exchange-Correlation Functional}
\begin{figure}[h]
\includegraphics[width=0.4\textwidth]{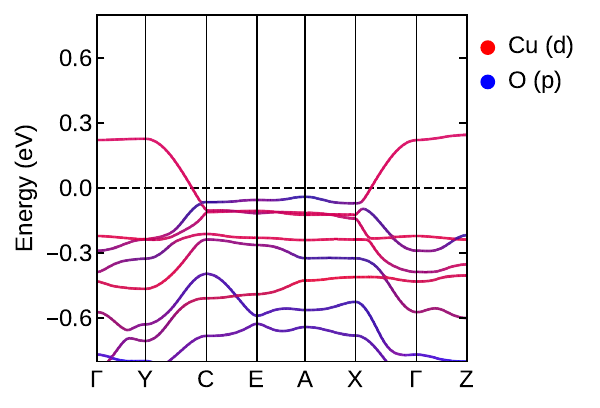}
\caption{Non-spin-polarized electronic band structure of \textit{Cu-2} calculated with r$^{2}$SCAN on PBEsol geometry. The Fermi level is set to 0 eV and is marked by the dashed line.}
\label{r2scan}
\end{figure}

We benchmarked our choice of exchange-correlation functional by comparing the DFT+U (U = 8~eV) magnetic and electronic structure properties with a select number of meta-GGA calculations using the r2SCAN functional. We find qualitatively similar results for the magnetic ground states regardless of functional choice, however the calculated exchange coupling strengths are greater with r2SCAN (see discussion in main text). We also calculate the electronic band structure for \textit{Cu-2} with r2SCAN finding, again, good qualitative agreement with very similar hole and electron pockets near the Fermi level. However, we find small shifts in the relative band filling at the Fermi level -- for instance the O-$p$ states are above $E_F$ with DFT+U, see Fig.~\ref{bands-pm}(c), while these are below $E_F$ with r2SCAN (Fig.~\ref{r2scan}). This suggests that correctly treating the O-$p$ states will also be important for accurately describing the low-energy physics of these systems.  

\begin{figure}
\includegraphics[width=0.48\textwidth]{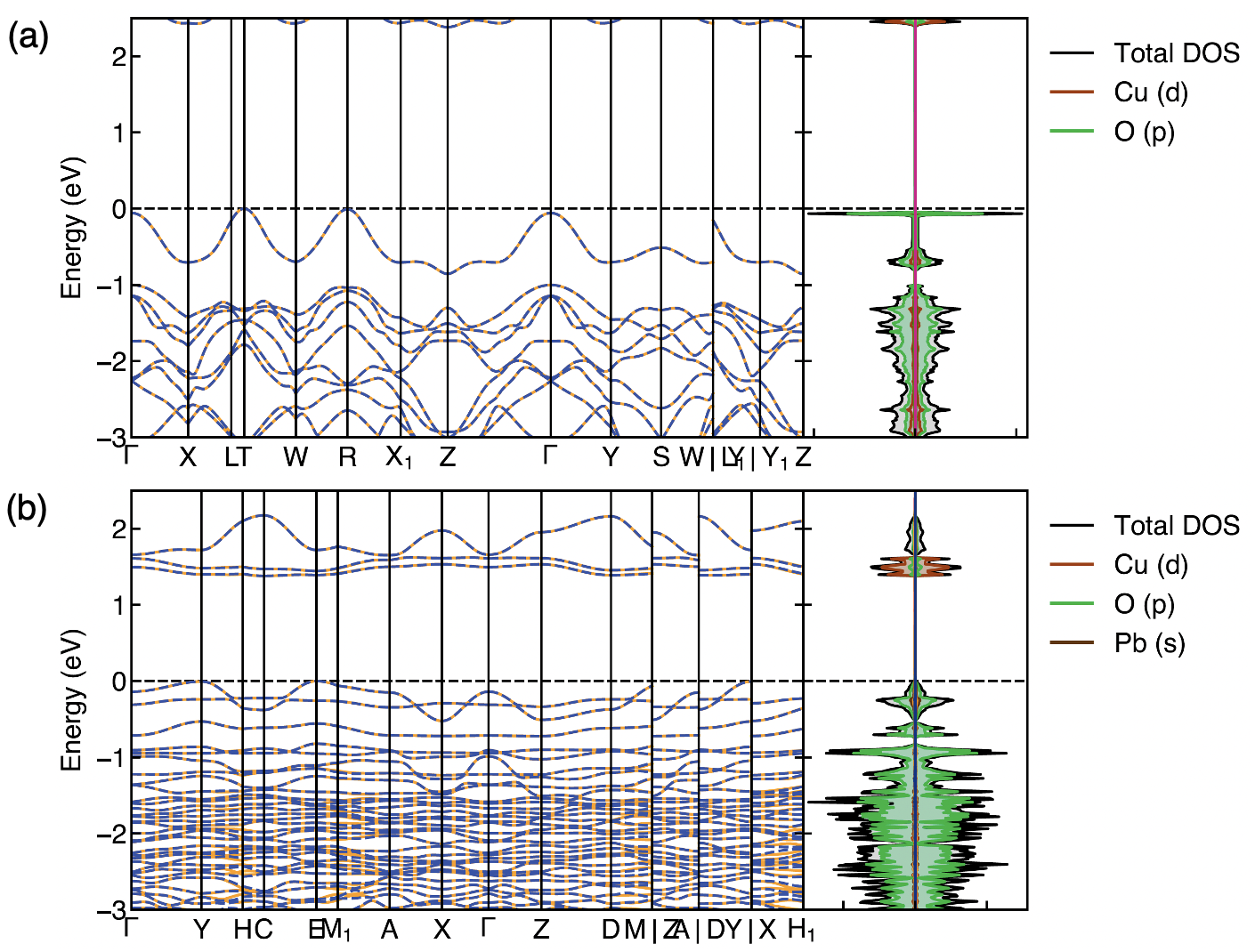}
\caption{Calculated spin-polarized electronic band structure (left) and corresponding density of states (right) for (a) Orthorhombic Li$_2$CuO$_2$, and (b) Cu$_4$Pb$_{6}$(PO$_{4}$)$_{6}$(O), \textit{`Cu-4'} where the Cu are all on the Pb-1 sites as depicted in Fig.\ref{structures}. The spin-up bands are depicted in solid orange, and the spin-down bands are dashed blue. The total density of states is shaded grey with projections shown of the Cu-$d$ orbitals (pink) and its neighboring O-$p$ orbitals (green). In both plots the Fermi level is set to 0 eV and is marked by the dashed line. } 
\label{bands-li-4cu-afm}
\end{figure}

\begin{figure*}
\includegraphics[width=0.95\textwidth]{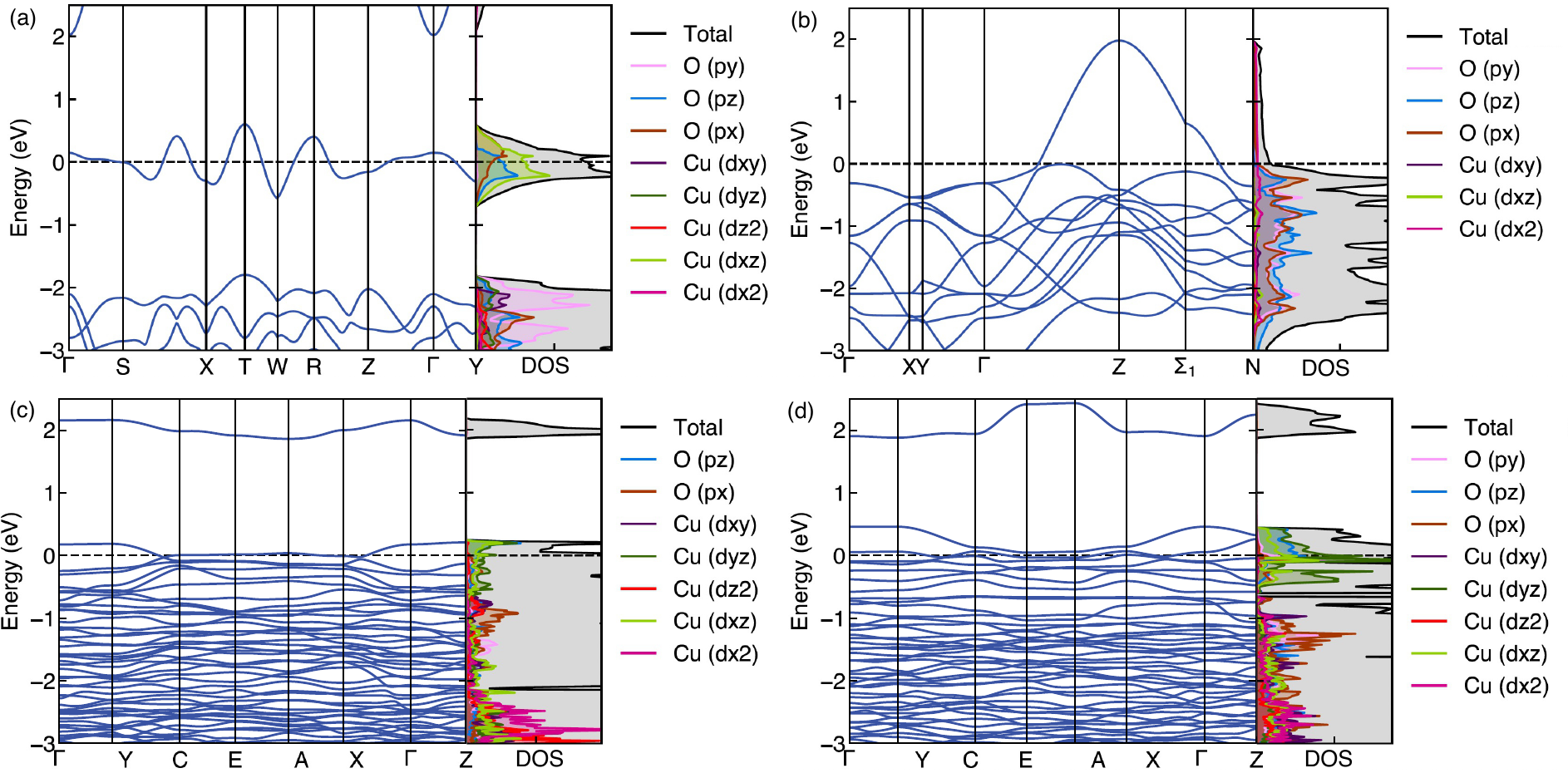}
\caption{Calculated non-spin-polarized electronic band structures (left) and orbital-projected densities of states (right) for (a) Li$_2$CuO$_2$, and (b) La$_2$CuO$_4$, (c) \textit{Cu-2} Cu$_2$Pb$_{8}$(PO$_{4}$)$_{6}$O, and (d)  \textit{Cu-4} Cu$_4$Pb$_{6}$(PO$_{4}$)$_{6}$O. The Fermi level is set to 0 eV in all plots and is marked by the dashed line.}
\label{pdos-comb}
\end{figure*}

\clearpage

\section{Structure Files}
The resulting relaxed structures in POSCAR format are given below. The specific geometry is given in the headers.

\begin{verbatim}
Cu-2 with oxygen vacancy; AFM 
    1.000000000000000     
     9.90522    0.73241    0.00000
    -6.06072    8.01068    0.00000
     0.00000    0.00000    6.95031
   Cu   Pb   P    O 
     2     8     6    25
Direct
  0.30643  0.71928  0.99956
  0.30643  0.71928  0.50044
  0.67644  0.32829  0.99414
  0.67644  0.32829  0.50586
  0.74168  0.70285  0.25000
  0.24732  0.26673  0.75000
  0.20552  0.99261  0.25000
  0.78750  0.99542  0.75000
  0.96639  0.19554  0.25000
  0.98062  0.77444  0.75000
  0.39661  0.40388  0.25000
  0.60204  0.62780  0.75000
  0.59361  0.01747  0.25000
  0.36831  0.95856  0.75000
  0.98601  0.57945  0.25000
  0.04285  0.41098  0.75000
  0.02251  0.01514  0.75000
  0.94150  0.93615  0.25000
  0.34740  0.51899  0.25000
  0.64208  0.49746  0.75000
  0.46666  0.83515  0.25000
  0.50027  0.13556  0.75000
  0.17308  0.65866  0.25000
  0.86526  0.36459  0.75000
  0.42389  0.56266  0.75000
  0.50784  0.10887  0.25000
  0.43392  0.84590  0.75000
  0.89636  0.39643  0.25000
  0.15998  0.59702  0.75000
  0.39417  0.31753  0.07069
  0.69093  0.74269  0.92464
  0.70577  0.07218  0.07152
  0.26185  0.90985  0.93319
  0.93494  0.63223  0.07137
  0.07539  0.34189  0.92949
  0.69093  0.74269  0.57536
  0.39417  0.31753  0.42931
  0.26185  0.90985  0.56681
  0.70577  0.07218  0.42848
  0.07539  0.34189  0.57051
  0.93494  0.63223  0.42863
\end{verbatim}
\newpage

\begin{verbatim}
Cu-4 with oxygen vacancy; AFM 
    1.000000000000000     
    10.26014   -0.19929    0.00000
    -7.09771    9.12119    0.00000
     0.00000    0.00000    5.99455
   Cu   Pb   P    O 
     4     6     6    25
  0.28115  0.71803  0.99746
  0.67207  0.25768  0.99870
  0.67207  0.25768  0.50130
  0.28115  0.71803  0.50254
  0.72766  0.67239  0.25000
  0.31283  0.32332  0.75000
  0.22296  0.03758  0.25000
  0.84480  0.01509  0.75000
  0.99048  0.21948  0.25000
  0.97660  0.79932  0.75000
  0.48079  0.37543  0.25000
  0.58274  0.67734  0.75000
  0.59368  0.01267  0.25000
  0.36291  0.96230  0.75000
  0.93906  0.55758  0.25000
  0.01974  0.42010  0.75000
  0.10394  0.06577  0.75000
  0.95491  0.87840  0.25000
  0.32658  0.36613  0.25000
  0.61026  0.56134  0.75000
  0.43884  0.84251  0.25000
  0.51620  0.13026  0.75000
  0.12783  0.61861  0.25000
  0.82971  0.35047  0.75000
  0.39966  0.61062  0.75000
  0.53425  0.11132  0.25000
  0.41442  0.85943  0.75000
  0.82080  0.37914  0.25000
  0.12972  0.59495  0.75000
  0.51237  0.32643  0.03597
  0.68259  0.78406  0.95699
  0.70447  0.06322  0.04214
  0.25132  0.91515  0.95880
  0.90062  0.60499  0.03995
  0.06023  0.37276  0.95959
  0.68259  0.78406  0.54301
  0.51237  0.32643  0.46403
  0.25132  0.91515  0.54120
  0.70447  0.06322  0.45786
  0.06023  0.37276  0.54041
  0.90062  0.60499  0.46005
\end{verbatim}

\end{document}